\documentclass[twocolumn,showpacs,prl]{revtex4-1}
\usepackage{amsmath}
\usepackage{graphicx}

\usepackage{color}
\usepackage{algorithm2e}

\begin{document}
\vspace*{0.35in}

\title{Transferable measurements of Heredity in models of the Origins of Life}
\author{Nicholas Guttenberg}
\author{Matthieu Laneuville}
\affiliation{Earth-Life Science Institute, Tokyo Institute of Technology, Tokyo, Japan}
\author{Melissa Ilardo}
\affiliation{Centre for GeoGenetics, Natural History Museum, University of Copenhagen, Copenhagen, Denmark}
\author{Nathanael Aubert-Kato}
\affiliation{Department of Information Science, Ochanomizu University, Tokyo, Japan and Leading Graduate School Promotion Center, Ochanomizu University, Tokyo, Japan }

\begin{abstract}
We propose a metric which can be used to compute the amount of heritable variation enabled by a given dynamical system. A distribution of selection pressures is used such that each pressure selects a particular fixed point via competitive exclusion in order to determine the corresponding distribution of potential fixed points in the population dynamics. This metric accurately detects the number of species present in artificially prepared test systems, and furthermore can correctly determine the number of heritable sets in clustered transition matrix models in which there are no clearly defined genomes. Finally, we apply our metric to the GARD model and show that it accurately reproduces prior measurements of the model's heritability.
\end{abstract}

\maketitle

\section*{Introduction}

Studies of the Origins of Life often come down to debate on the proper definition of 'life'. However, underlying that philosophical point are a number of distinct phenomena which are exhibited by modern biological life whose emergence must be explained. By studying these particular, quantifyable phenomena, it is possible to make progress in our understanding even if the overarching philosophical question remains murky. For example, by focusing on 'replication' in detail there have been many advances in understanding autocatalytic chemistry, both experimentally and theoretically\cite{kauffman1986autocatalytic, orgel2000self, hordijk2004detecting, saladino2012formamide, england2013statistical, cronin2014evolution}, even if those systems may not fully qualify as life. 

One of the milestones of a theory of the Origins of Life is the understanding of how to bridge the gap between abiotic conditions and the onset of evolution. Along with replication, the mechanics of Darwinian evolution require that the system be capable of heritable variation and also be subject to selective pressures. Selective pressures are readily available --- chemical reaction rates are in general very sensitive to environmental parameters --- but it is more difficult to obtain the ability to inherit variations. In order to understand how chemical systems become capable of evolution we must then ask: what are the necessary and sufficient elements for a system to be capable of heritable variation? 

Because this question is very broad, it is well-suited to abstract models, in which the dynamics of the system may be simplified in order to determine what elements are truly necessary. There are several examples of models which appear to give rise to forms of heredity\cite{fontana1994arrival, markovitch2012excess, vasas2012evolution, virgo2013autocatalysis}. One of the difficulties in working with abstract models of something as broad as the Origins of Life, however, is that it is often unclear how to compare the results of these different models because they use very different underlying metaphors --- chemical reaction networks, mathematical functions, combinatoric assemblages of materials, and cellular automata among others. In order to help connect those abstract efforts to the more concrete goals of understanding the Origins of Life, we need devise a metric which can be applied very generally to a wide array of abstract models. Furthermore, we would like this metric to be equally applicable to experimental systems, so that the abstract models can be evaluated to see whether or not they are truly predictive.

Heritable variation in modern biological systems proceeds from a well-defined information-carying molecule and well-separated individuals which comprise a population. In this case one can conceive of a very clear way to understand heredity by directly comparing the DNA of an organism to its parent. This leads to understandings of mutation, the process of fixation, and also a way to understand natural selection as a population-level effect. In asking questions about the emergence of life, however, it is likely that we will need to deal with intermediate systems in which things are not so clearly separated. We may not know what the information-carrying component, or the information may be distributed across all the degrees of freedom of the system. In addition, if asking questions about pre-cellular life, there will not necessarily be a way to distinguish individuals in the population. We may also not know the timescale of a single generation of replication or be working with systems in which the dynamics are more continuous and distinct generations cannot be defined.

In order to deal with this, we attempt to find a definition of heritable variability that does not depend on there being a known information-carrying molecule, well-defined individuals, or a known timescale associated with the turnover of generations. As such, we will not refer directly to the idea of replication with a particular fidelity, but instead try to look at the long-time behavior of dynamical systems. Heredity is then related to the property of history-dependence. If a system's state remains dependent on its history at long times, that enables it to retain and remember information. For example, a system with bistability is capable of remembering a single bit of information. 

What distinguishes heredity in particular from any form of memory is that it is memory which is amplified extensively via replication. A system with heredity, as opposed to one with just memory, can contain multiple copies of one or more pieces of information, and each of those pieces of information is constantly being copied. This distinction is what allows systems with heredity to undergo evolution via population dynamics. The consequence of this is that even if the dynamics of a piece of the system would be bistable or multistable, if the various pieces of the system are amplified at different rates then at long times and for large systems the fastest growing state will dominate. In population genetics this is the phenomenon of competitive exclusion\cite{hardin1960competitive}. As a result, the number of heritable states we detect will be dependent on how long we measure for --- in other words, because we do not a-priori know the appropriate choice of timescale, we cannot recover a unique measurement of how much heritable variation is possible.  Different investigators with different systems would need to appropriately choose that timescale for their experiments, and there may not be a good way to say what that timescale should in fact be. This makes it hard to compare results across models.

If however one performed the above experiment in a variety of different environments (selection pressures), each environment would potentially pick out a different dominant state at long times. This is still a somewhat arbitrary choice, as it is necessary to pick a particular distribution of environments tuned to the given system, such that the perturbation is strong enough to re-order the competing states but not so strong as to completely alter their structure. However, there is an advantage to this approach, in that it separates out two distinct mechanisms for the system having a different steady-state configuration. One mechanism is that as the environment is perturbed, there is a corresponding perturbation to the location of the long-time stable state. That is, when a particular reaction rate is altered infinitessimally, the resting concentration of each compound is also altered infinitessimally. The other mechanism is that if there are a number of states that are local maxima in the replication rate, an infinitessimal change in the replication rates can alter which state is the global maximum. This means that a small change in selection pressure may give rise to a discontinuous large change in the long-term dominant state. This second mechanism is what lets us count the number of local maxima even if the dynamics only ever finds the global maximum. Since the two mechanisms behave qualitatively differently under a distribution of selection pressures, we can focus on distinguishing the different types of variation, and by doing so the results should be relatively insensitive to the particular choice of selection pressures (so long as the distribution is wide enough to find all the local maxima).

Therefore, the central idea of our metric of heredity is that what we are measuring has to do with the set of ways in which a given dynamical system can respond to a distribution of selective pressures. The distribution of selection pressures connects with all of the different possible variable degrees of freedom of the system, but due to the presence of replication the results of this are projected onto the subset of those variable degrees of freedom which are also heritable. Essentially, we can measure the 'genotype' hidden behind the set of phenotypes we observe, by seeing correlations between phenotypes which persist over a variety of different selection pressures (including pressures which explicitly differentiate between those phenotypes).

What remains is to devise a computational method to evaluate how many different genotypes are possible, given a set of outcomes. In general, real systems will not only have fluctuations which may blur the distinction between different species, but also will not have a clear separation between parts of the system which are responsible for the mechanics of replication and parts of the system which are responsible for holding heritable information. An autocatalytic chemical system, for example, will be accompanied by the decomposition byproducts of all of the members of the autocatalytic cycle. Those decomposition byproducts may themselves vary in relative concentration depending on the particular kind of selection pressure that is applied, even within the same 'species' of core autocatalytic cycle.

In general, we will not be able to guarantee that any algorithm we find will perfectly identify the variation between heritable states versus the variation within a single heritable state. This problem is nothing new --- in bioinformatics, it is often difficult to precisely define a bacterial species, and instead a clustering classification called 'organizational taxonomic unit' must be used. What we can do is to try to characterize the behavior of the metric when given ambiguous cases and understand the failure modes. As such, we will have some ability to devise tests to recognize when the metric cannot be applied accurately and to provide bounds on the error. 

First we will explain the algorithm for computing our heredity metric. Then, we will apply this metric to an artificial data set with a known number of species, distributed in different ways and with different mutation rates, in order to show that it can correctly determine the number of species present and also to show what happens when it breaks down. Next, we will examine another example system in which we construct a transition matrix describing a set of overlapping autocatalytic networks, and show that the metric can accurately detect the number of modules. We will then apply the metric to the GARD model, a non-trivial model system  that has a distributed, 'compositional' heredity which has been extensively studied\cite{segre1998graded, segre2000compositional, segre2001lipid, vasas2010lack, markovitch2012excess}. 

\section*{Methods}

Our algorithm for detecting potentially heritable states is as follows, taking as input the data matrix $\textbf{P}$ and giving as output the number of heritable states of the dynamics $N_S$. Each row of the data matrix is a single observation of the dynamical system of interest at long times, driven by a particular selection pressure (where each row corresponds to a different selection pressure), and each column corresponds to a feature of the system --- these can be binary features or scalars which have been normalized with respect to eachother to have similar variance.

\SetKw{Optional}{Optional}
\SetInd{1em}{1em}
\begin{algorithm}
\KwData{Matrix of observations $\textbf{P}$ taken at }
\KwResult{Number of heritable states $N_S$}
\Begin{
Subtract the mean from each column of \textbf{P}\;
Add copies of data rows for regularization\;
Principal Component Analysis of $\textbf{P}$ $\rightarrow$ $N_{eigs}$ eigenvalues $\lambda_i$ in descending order\;
Set the index $j = N_{eigs}$\;
\Repeat{$j$ not yet converged}{ 
 Compute $\lambda_> = \frac{1}{j} \sum_{i=j}^{2j} \lambda_i$ ($\lambda_i=0$ if $i>N_{eigs}$)\;
 Set the threshold $T=\alpha (\lambda_1 - \lambda_>) + \lambda_>$\;
 Set $j$ to be the index of the first eigenvalue whose value is less than $T$\;
}
The measured number of heritable states is $N_S=j+1$\;}
\end{algorithm}

As part of the development of this metric, we tried a number of different techniques: recursive feature elimination to find the subset of minimally correlated features, information-theoretic measures such as sequence entropy and mutual information, clustering algorithms (K-means and Agglomerative Clustering), and dimension-reduction techniques (PCA). There are many methods that all seem feasible, and we do not have a strong argument to pick one above all the others on the grounds of first-principles. Instead, we present the method which out of our test set had the best performance both in terms of robustness, ability to detect different heritable states, and failure modes which are least confusing. Out of the techniques we tried, we found the best results using PCA combined with an analysis of the eigenvalue spectrum, and so we focus on that.

The reasoning behind using PCA to detect the number of heritable states is that if there are $N_S$ different heritable states distributed in a space of much higher dimensionality, we expect that there will not be collinearity between any pairs of states in the case where there is any randomness involved in the location of the states in the high-dimensional space. As such, a set of $N_S$ points defines a subspace of dimension $N_S-1$, which is what the PCA is detecting. If fewer components are retained, then one heritable state must be explained as a superposition of the others, but as long as the space of features is of much higher dimension than the number of different heritable states, this projection will generally have an error whose magnitude comes from the underlying distribution of the heritable states of the system rather than coming from the fluctuations. By using PCA and looking at the eigenvalue spectrum, we attempt to detect that difference. 

The input to the PCA is a matrix of data $\textbf{P}$, for which each column is some observable property of the system se (features) and where each row is the final system state under a particular randomly sampled selection pressure. These features are things such as whether a particular chemical is present beyond a certain threshold concentration, whether a certain gene is present or absent, etc --- they must be chosen by the investigator appropriately to the model in question. In general, because of the properties of PCA, features should be normalized with respect to each-other and have a zero mean over the distribution of the data.

The PCA then finds a set of new mutually-independent features which are linear combinations of the given ones such that the covariance between features is zero. These features are sorted with respect to the amount of variance in the data associated with each feature --- this information is contained within the eigenvalues. If there are certain directions which encode the differences between heritable states, the variance associated with those directions will grow as more features are added. This is because that variance comes from the distance between heritable states in the feature space. As one increases the dimensionality of that space, distances within that space will grow monotonically, as each added dimension corresponds to an additional strictly-positive contribution to the total distance. 

At the same time, the per-eigenvalue variance that is due to random fluctuations (e.g. uncorrelated with selective pressures) will remain roughly constant because as new random variables are added, new eigenvalues are also added at the same rate. So as long as there are a sufficient number of features, we expect that we should be able to use the eigenvalue spectrum to distinguish those that come from the system being in globally distinct states and those that come from noise. Similarly, as more data is added, we expect the algorithm to converge. In general, assuming a sufficient number of features, the set of heritable states is a smaller-dimensional subspace than the set of possible random fluctuations. As such, the distribution of the data will converge more quickly in directions corresponding to heritable states than it will in the directions corresponding to fluctuations, and so we expect an increasing contrast as we add more data.

Once there are sufficient features and data points, we can plot the eigenvalues sorted by rank. In cases where there are multiple attractors for the system state, these show up as a number of large eigenvalues. There is then usually a large drop, followed by a large number of small eigenvalues associated with the fluctuations of the system around these attractors. Some example eigenvalue plots are shown in Fig.~\ref{TransitionTestCase}. While the gap can often be located by eye, we need the algorithm to reliably detect it in an automated way. For this purpose, we use an iterative procedure to localize the gap where there is a significant separation between the set of eigenvalues before the break and the set of eigenvalues after the break. 

In general, the eigenvalue spectrum from PCA on the expected kind of data has a quickly-decaying part of fixed length (from heritable variation) followed by a slowly-decaying tail whose length depends on the number of features and samples. By doing a local average of the eigenvalues below the cutoff, it is possible to detect the relevant height of the noise floor generated by mutation and the like in order to subtract it out. Roughly speaking, this procedure finds the point at which the local derivative begins to be significantly steeper than the average slope from the origin to that point --- essentially, its a form of cliff-detection that normalizes with respect to a background mean and mean slope. The parameter $\alpha$ controls the sensitivity of the algorithm. If $\alpha$ is set to be small, then in general the algorithm will detect finer differences between heritable states; however, when there are many heritable states, the error tends to be in the form of large over-estimation of the number of states in the system. If $\alpha$ is made larger, then the algorithm tends to saturate and cannot detect more than a certain number of heritable states, but errors are made in the form of under-estimation. 

In general, by examining the convergence pattern of the algorithm for a few cases (as in Fig.~\ref{SpeciesConvergencePlot}), it is possible to find an optimum value for $\alpha$ which produces the most stable results with as little data as possible. We recommend that $\alpha=0.1$ be taken as a good initial parameter based on the results of our test cases. Sensitivity to the choice of $\alpha$ may be an indication that the algorithm is failing to detect the difference between heritable variation and the baseline fluctuations, and so it is generally a good idea to examine a convergence plot to ensure that the algorithm is behaving consistently for whatever choice of $\alpha$ ends up being used.

We have made a simple Python script using Scikit-Learn\cite{pedregosa2011scikit} which implements this metric given a file containing features and observations available at \url{http://www.github.com/ModelingOriginsofLife/Heredity} so that other researchers can easily apply it to their own data sets.

\subsection*{Consistency checks}

Our metric works best when the system has a small number of heritable states compared to the number of features and independent measurements provided. For a small number of heritable states, the gaps are generally very well-defined, but as the number increases it becomes harder to distinguish them from the noise. The algorithm strictly cannot detect more heritable states than there are features. For this reason, as many potentially relevant features as possible should be provided in the data. Because of this limit, systems with combinatorically large genetic spaces must be broken up before they can be analyzed with our metric. 

This can be done by controlling the distribution of selection pressures. Our metric only detects heritable states which are influenced by variation in the selection pressure. If one constrains the selection pressure to a low number of dimensions, that will isolate a particular subset of heritable states. One can then use different selection pressure distributions to find different subsets of heritable states in the same system. This is analogous to measuring the heritability of a single gene at a time, rather than looking at all possible genes at once.

When directly applying selection pressure to a particular subset of features, the variance in those features will scale differently than the variance coming from underlying noise. This can create an artificial signal that appears to correspond to a set of heritable states where really all that it is detecting is the distribution of external impulses. As such, it is best to exclude any features which are being directly driven by the externally varying selection pressure from the analysis. For example, in a chemical system with a number of compounds one form of restricted selection pressure would be to add reactions which decompose certain combinations of compounds. The compounds which can be directly decomposed by this reaction should not be included as features in the analysis.

Another possible solution is to consider the properties of systems not in terms of a fixed count of the number of heritable states, but in terms of scaling laws. If we were to imagine applying our metric to DNA sequences of length $L$ being replicated via the polymerase chain reaction, then the number of heritable states is expected to scale as $\exp(L)$. If $L$ is large, we cannot expect to sample all possibilities, but if $L$ is small then it is still possible to do so. As such, we could imagine fixing $L$ (or filtering the results according to $L$) and then performing the analysis for different $L$ values. In the DNA system, we might be able to see that the number of heritable states is scaling in an open-ended way as we increase $L$ (so the heredity is unlimited), and that it is growing exponentially with $L$ (so the heredity is combinatoric in nature). On the other hand, if we were to do something similar in other dynamical systems, we might see a characteristically different scaling. 

There may also be problems when there is no heredity at all in the system. When this happens, all directions have the same apparent variance and so the behavior of methods to detect the 'jump' between sets of eigenvalues often becomes ill-defined. This particular error can be detected by examining the convergence curve of the algorithm periodically as one accumulates more data rows --- if there is only one species, the number of detected species will keep increasing as one adds data, until it is equal to roughly 30\% of the number of features. 

It is also possible to use a form of regularization to help prevent this error in the first place. To do so, we artificially add a small number of data rows with very low variance (essentially an artificial 'species'), and then subtract one from the detected number of heritable states at the end of the analysis. In our examples, we do this by taking a random data row from the system and making $10$ copies of it. This guarantees some degree of consistency with the underlying statistics of the data (which might be violated if one were to use a completely arbitrary artificial species). We find that this does not significantly harm convergence elsewhere, but strongly helps in the case where there is actually only one heritable state in the system.

\section*{Results}

We present a number of test cases in order to evaluate the performance of our metric. We start with a straightforward case in which we define apriori the structure of the heritable states, and see if our metric can detect the number of heritable states built into the system. In the second case, we model the dynamics of mutation and replication with a transition matrix and evaluate the algorithm on the resulting attractors --- this lets us test whether the method of using a distribution of selection pressures works for picking out heritable states. The third test case is the graded autocatalysis replication domain (GARD) model\cite{segre2000compositional,markovitch2012excess} of compositional heredity in replicating vesicles, which lets us test our metric on a non-trivial system whose heredity properties have been studied elsewhere.

\subsection*{Species Detection}

In this test case, we artificially prepare a population of distinct heritable states in order to see whether or not the metric properly detects how many states are present. Here, we do not try to see heritability emerge from an underlying dynamic but instead impose it directly in a population genetics framework such that each heritable state is the equivalent of a distinct species of organism. We implicitly allow the process of competitive exclusion proceed to completion in each case, so that the final state of the system for each independent run is a particular species picked out of the set of possible species, plus mutation.

A system state is comprised of a number $N_f$ of binary features. Each species is a particular system state, with a fraction $f_1$ of bases set to $1$, and the rest set to zero. When generating a novel end-state, we pick a random species and then apply mutation to its binary feature vector --- each feature has a probability $m$ of being flipped. We generate a population of $N_P$ such states, which we then analyze using our metric.

\begin{figure}
 \includegraphics[width=\columnwidth]{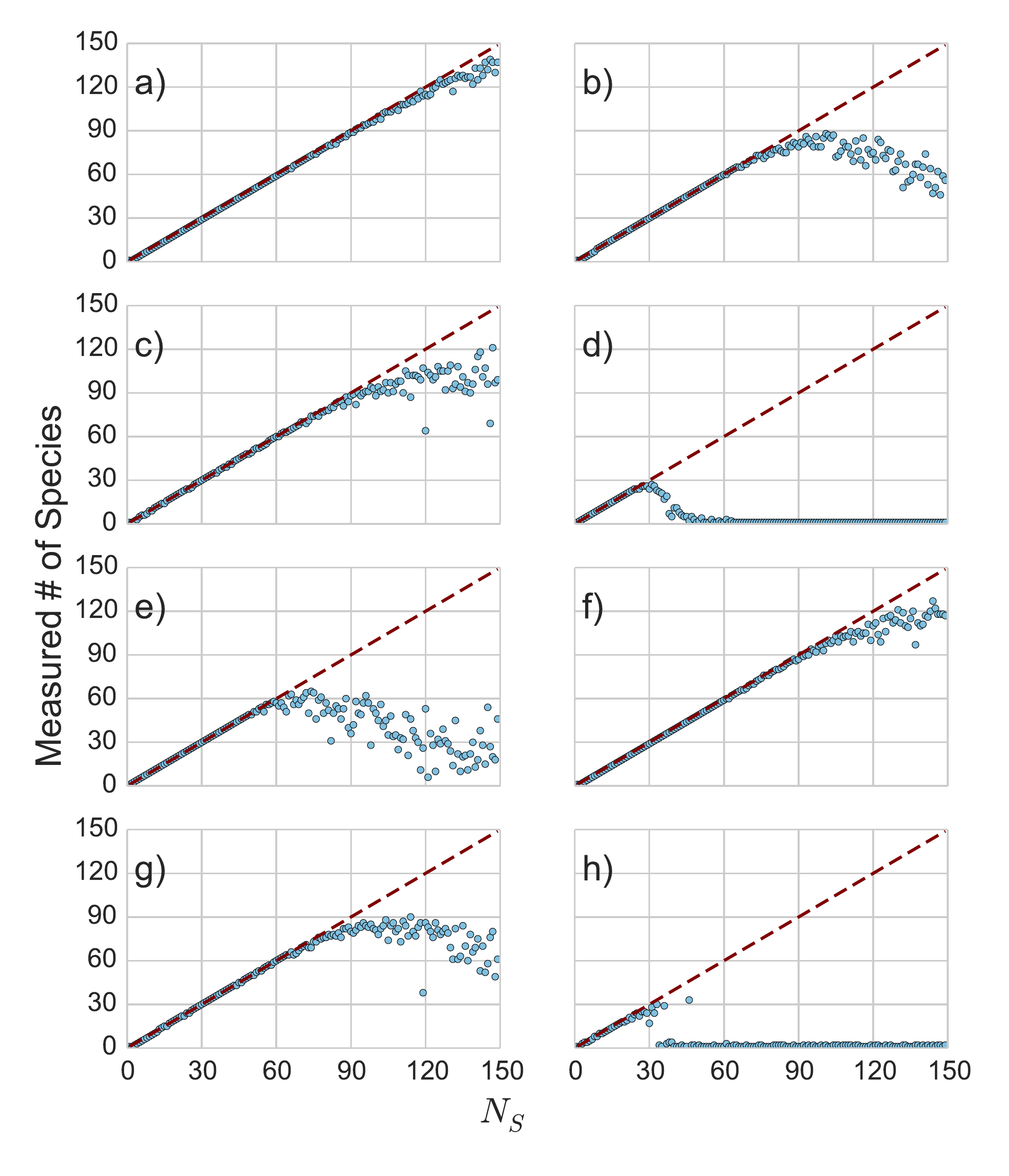}
 \caption{\label{SpeciesPerformance}{\bf Behavior of the heredity metric on the species detection test case.} For these plots, perfect detection corresponds to the line $y=x$. These cases use $\alpha=0.1$, $N_P=1000$, $N_f=1000$, and 10 extra duplicate sequences for regularization. a) $f_1 = 0.5, m=0.05$, b) $f_1 = 0.5, m=0.2$, c) $f_1 = 0.05, m=0.05$, d) $f_1 = 0.05, m=0.2$, e) $f_1 = 0.05, m = 0.05,0.2$, f) $f_1 = 0.05, 0.5, m=0.05$, g) $f_1 = 0.05, 0.5, m = 0.05, 0.2$, h) $f_1 = 0.05, m=0.05$, heterogeneous population}
\end{figure}

We discuss a number of particular configurations of this process, in order to try to see how the algorithm responds to various possible problems that might arise in real data. The base case we will consider has $f_1=0.5$, $m=0.05$, $N_f=1000$, $N_P=1000$, and a variable number of species (sampled uniformly). This is a relatively gentle case, as the average distance between the binary vectors for two different species is $d \approx 22$, whereas the average distance between members of the same species is $d \approx 10$, and so the various heritable states should be well-clustered. The performance of the algorithm in this case is shown in Fig.~\ref{SpeciesPerformance}a. In this base case, the correct number of species is detected to within one species  error out to $N_S = 150$ --- this means that in many cases there may only be a few individuals of those species in the system.

\begin{figure}
 \includegraphics[width=\columnwidth]{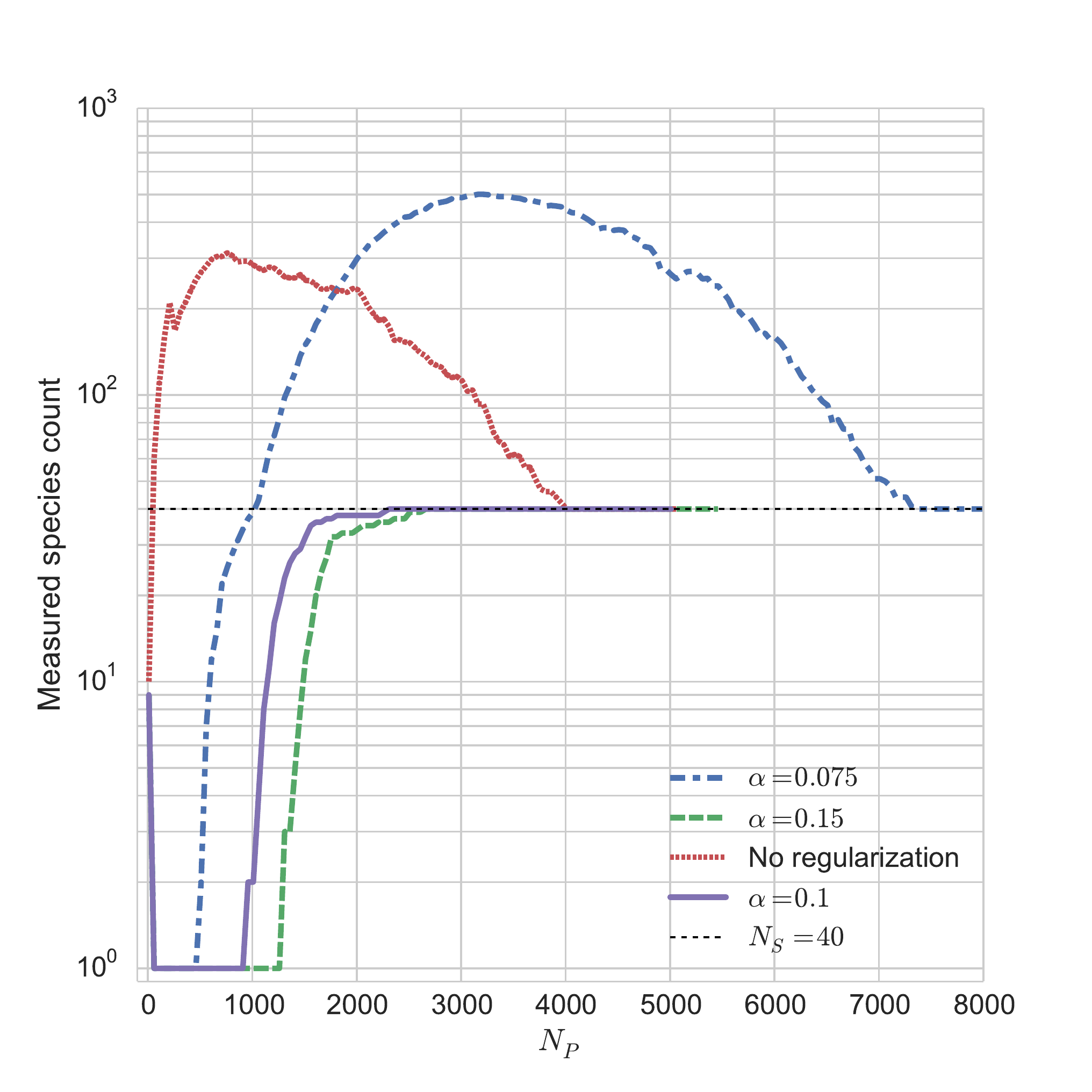}
 \caption{\label{SpeciesConvergencePlot}{\bf Convergence of the measured number of heritable states as the number of data points is increased.} This shows the effect of regularization and of the choice of $\alpha$ on the convergence pattern in this particular test case. The parameters for these data are $f_1=0.05, m=0.2, N_S=40$. }
\end{figure}

We may also want to measure systems in which fluctuations are very large compared to the systematically inherited variance, and so we want to see what happens when the mutation rate $m$ becomes large in order to see how the algorithm fails. Our second configuration is the same as the base configuration, but with a much higher mutation rate: $m=0.2$. This means that the average distance within a species is $d \approx 19$ --- close to the average distance between a random pair of species. As the number of species increases, we expect to have several of the random species end up being closer together than the size of their mutational haloes. The results are shown in Fig.~\ref{SpeciesPerformance}b. For small species counts, the algorithm is still accurate, but at larger species counts there is a point at which the algorithm fails to detect all of the species present.  An increased $N_P$, corresponding to the availability of more data samples, allows the algorithm to detect the correct number of species. As a result, a convergence plot (Fig.~\ref{SpeciesConvergencePlot}) showing the measured number of species calculated as a function of the number of data rows $N_P$ may be useful in determining the trustworthiness of the metric for a given system.

Another common complication is that the distribution of values of the binary features is not uniform. If for example there is a particular chemical produced in only one of the heritable states, then a feature associated with that chemical would be zero most of the time but rarely would be one. This would not simply be an improbable fluctuation, but instead signifies the presence of a particular heritable state. In terms of this test case, the consequence of these rare but significant feature values is that the average distance between different species is reduced. We look at two situations here: one in which $f_1 = 0.05$ and $m=0.05$ (Fig.~\ref{SpeciesPerformance}c) and one in which $f_1 = 0.05$ and $m=0.2$ (Fig.~\ref{SpeciesPerformance}d). This effect does somewhat destabilize the algorithm (especially in the high mutation case), but in general is less harmful than simply having a high mutation rate, even though for $f_1 = 0.05$ and $m=0.05$ the average distance between species is the same as the average distance between members of a species.

The next configurations we look at is when the binary features are not homogeneous. We look at three examples of this: one in which half of the features have $f_1 = 0.5$ and the other half have $f_1 = 0.05$ (Fig.~\ref{SpeciesPerformance}e); one in which half of the features have $m = 0.05$ and the other half have $m = 0.2$ (Fig.~\ref{SpeciesPerformance}f); and one in which we have four types of features corresponding to all combinations of the former cases (e.g. $m = 0.05,2$ and $f_1=0.05,0.5$ (Fig~.\ref{SpeciesPerformance}g). In general, the algorithm performs reasonably well for these cases, tending to make errors in the direction of under-estimating the number of species. 

The final case we look at is when the species themselves are not uniformly sampled. Instead, we sample the various species with probability such that:

\[
p_i =
\begin{cases}
(1-1/N_S)^{i-1}/N_S & i\leq N_S-1 \\
(1-1/N_S)^{N_S-1} & i = N_S 
\end{cases}
\]

This means that the population ratio between the most common and least common species grows as $N_S-1$. Even with a low mutation rate ($m = 0.05$), this appears to make the problem significantly harder when the number of species to distinguish grows. This makes sense, as at some point the rarest species only have a handful of examples in the population and are comparable to variation due to mutational noise. As might be expected, the algorithm degrades in the form of under-detecting the correct number of species. One complication however is that as the (apparent) mutational floor grows, this can interfere with the detection of species that were correctly detected previously. This case is plotted in Fig.~\ref{SpeciesPerformance}h.

\begin{figure}
 \includegraphics[width=\columnwidth]{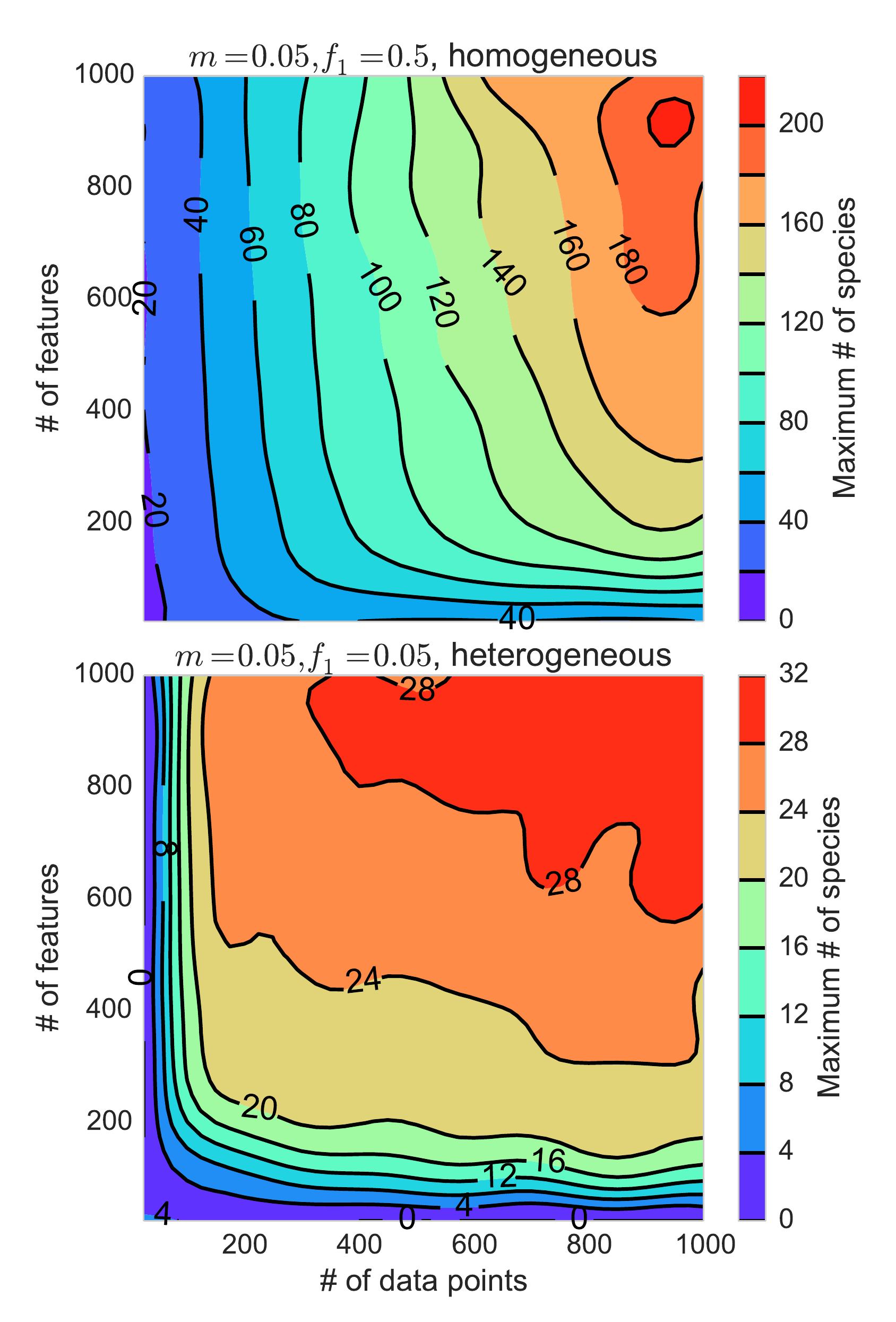}
 \caption{\label{Confidence}{\bf Confidence regions for detecting different numbers of heritable states.} Upper: detection confidence regions (50\% error) for the easy case ($m=0.05$, $f_1 = 0.5$, uniform distribution of species). In this case the algorithm is mostly data-limited, accurately detecting at most a number of species roughly equal to $0.25 N_P$. Lower: detection confidence regions for a harder case ($m=0.05$, $f_1=0.05$, heterogeneous distribution of species). In this regime the algorithm is feature-limited --- more features are needed to detect more species.  }
\end{figure}

When applying this algorithm to a novel system in which the behavior is not already well-understood, it is important to recognize potential signs that the algorithm may not be detecting the correct number of heritable states due to an insufficient number of features or data points. To this end, we varied the number of features and data rows in this simple case and measured the number of species that can be included before the algorithm's performance degrades. Specifically, we look at the point at which the error in the measured number of species exceeds 50\%. The results of this are plotted for an easy test case ($f_1 = 0.5, m = 0.05$, homogeneous population) and a difficult test case ($f_1 = 0.05, m = 0.05$, heterogeneous population) in Fig.~\ref{Confidence} (upper and lower panels respectively). In the case of a homogeneous population, the ability of the algorithm to resolve distinct species appears to be linear in the number of data points (requiring roughly $4 N_S$ data points in the easy case) and sublinear in the number of features (though the algorithm can never detect more species than there are features due to the properties of PCA). When the population is heterogeneous, then the scaling with the number of data rows seems to be more severely limited, dependent on the functional form of the distribution of heritable states.

\subsection*{Transition Matrix Model}

In the previous case, we artificially constructed a specific distribution of species. In real cases, however, we expect to use a distribution of selection pressures in order to probe the heritable states of a system. We extend the idea of the former test case to take into account dynamical processes and the response to selection by using a transition matrix style approach to a replicative process. In this model, the overall system is described by a distribution of random walkers over a set of nodes. There is a transition matrix which governs the movement of the random walkers between nodes. Additionally, replication is implemented by way of a per-node replication rate, such that each iteration more walkers may leave a node than initially entered it.  
This can be thought of as writing a transition matrix where the columns are not normalized to $1$.

This sort of model is similar to quasi-species in population genetics. In such systems, there is an error-threshold\cite{eigen1988molecular} with respect to the balance between mutation (links from a node to other nodes) and replication (links from a node to itself). When the selective advantage of a node is sufficiently strong compared to mutational diffusion, then that node ends up forming the core of a well-defined species --- e.g. as one increases the number of potential nodes, the ratio between the average population of the nodes belonging to the species and the average population of the nodes not belonging to the species diverges to infinity. On the other hand, if mutation is stronger than replication then this ratio becomes independent of the total number of nodes (e.g. the population fills the space of possibilities semi-uniformly). 

In our case, we would like to build a transition matrix such that its' overall structure allows for the existence of clusters of nodes which mutually exceed the error threshold. This is a closer analogy to autocatalytic chemical systems, in which a given compound will not necessarily directly replicate itself, but will instead proceed through a number of intermediaries (which may have their own side-reactions, or be part of multiple autocatalytic cycles). 

Such transition matrix based systems are linear, which means that the dynamics can be solved directly by computing the eigenvalues of the transition matrix. At long times, the state of the system will always be dominated by the largest eigenvalue, which will tend to not be degenerate unless there is a strong symmetry in the transition matrix (e.g. something like a block-diagonal structure). This is equivalent to competitive exclusion. As such, even if we build a transition matrix with multiple species which replicate with fidelity, at infinite time we should only observe a single system-wide state corresponding to the largest eigenvalue. This means that if we wish to detect the cluster structure of the transition matrix strictly from state data, we cannot do so unless we apply a distribution of selection pressures (which takes the form of permutations to the transition matrix) or otherwise stochastically drive the system strongly enough to overcome the gaps between its large eigenvalues.

To produce this type of topology, we use the following procedure:

\begin{enumerate}
 \item Randomly assign each node membership to a particular cluster $k_i$ (out of $N_C$ total clusters), and a replication rate $r_i = r_0 + \delta r \eta_i$, where $\eta_i$ is a random variable distributed uniformly between $[-1,1]$
 \item For each node, create $N_L$ random links to other nodes. If the same pair of nodes is chosen multiple times, sum the weights of the links
 \item For each such link, assign a weight: $(1+\gamma)/2$ if the nodes belong to the same cluster, or $(1-\gamma)/2$ if the nodes belong to different clusters.
 \item Normalize the weights of links leaving each node so that they sum to $r_i$
\end{enumerate}

\begin{figure}
 \includegraphics[width=\columnwidth]{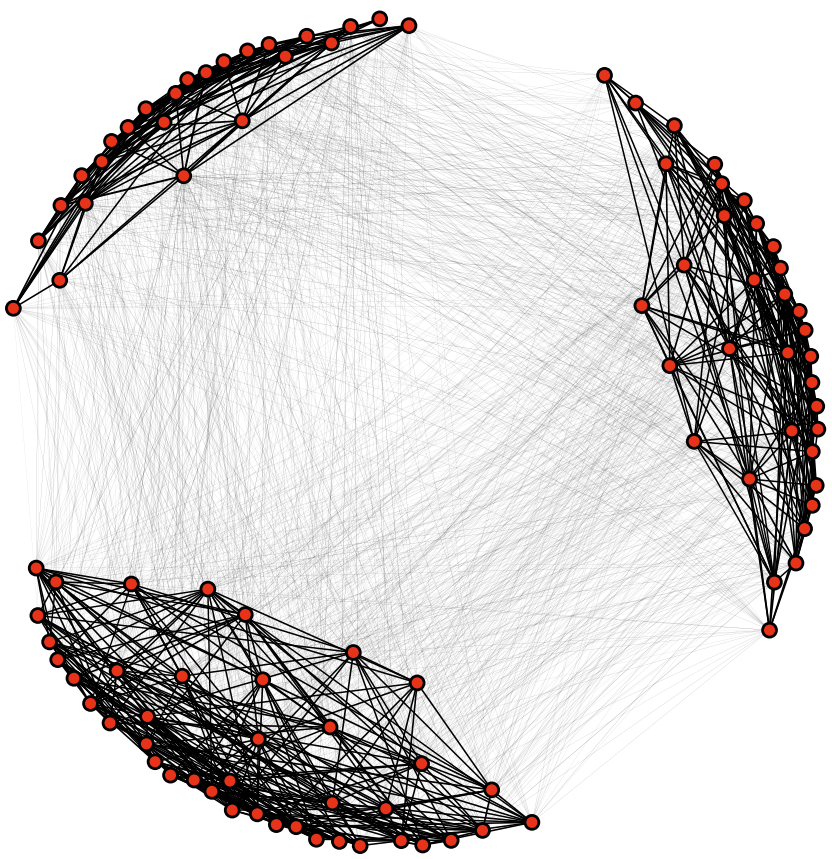}
 \caption{\label{NetworkStructure}{\bf Transition matrix network structure.} Example structure of the transition matrix with $100$ nodes, $N_L=20$, and $N_C=3$ visualized as a network. The darkness of edges is proportional to the probability of a transition along that edge. Even though there are more between-cluster links than within-cluster links, within-cluster links are a factor of $1000$ more likely to be followed than between-cluster links, meaning that overall a random walker tends to stay within its current cluster.}
\end{figure}

This produces matrices with clusters that still have weak inter-cluster connections. An example of such a network with $100$ nodes, $N_C=3$ clusters, and $N_L=20$ links per node is shown in Fig.~\ref{NetworkStructure}. We generally have each node connect to $20\%$ of the other nodes in order to ensure that there are few nodes which do not link to any other member of their cluster (which becomes a factor as the number of clusters grows).

Given a transition matrix $\textbf{T}$ of this form, we can attempt to apply our metric by simulating the dynamics of this system. Rather than simulate individual random walkers, we work with a vector encoding their population. At each iteration, we apply the transition matrix to this vector $p_{t+1} = \textbf{T} p_t$ and then normalize the vector to $1$ (this has no effect on the dynamics, but helps keep the numerics stable). Each run, we start with a random population vector and create a permuted transition matrix by randomly reducing the on-site replication rates of half of the nodes by about $1\%$ in order to create a distribution of selection pressures. In addition, we allow the replication rates at each node to vary randomly (uniformly) by $\pm 1\%$ each iteration to represent the effects of a noisy environment. 

\begin{figure}
 \includegraphics[width=\columnwidth]{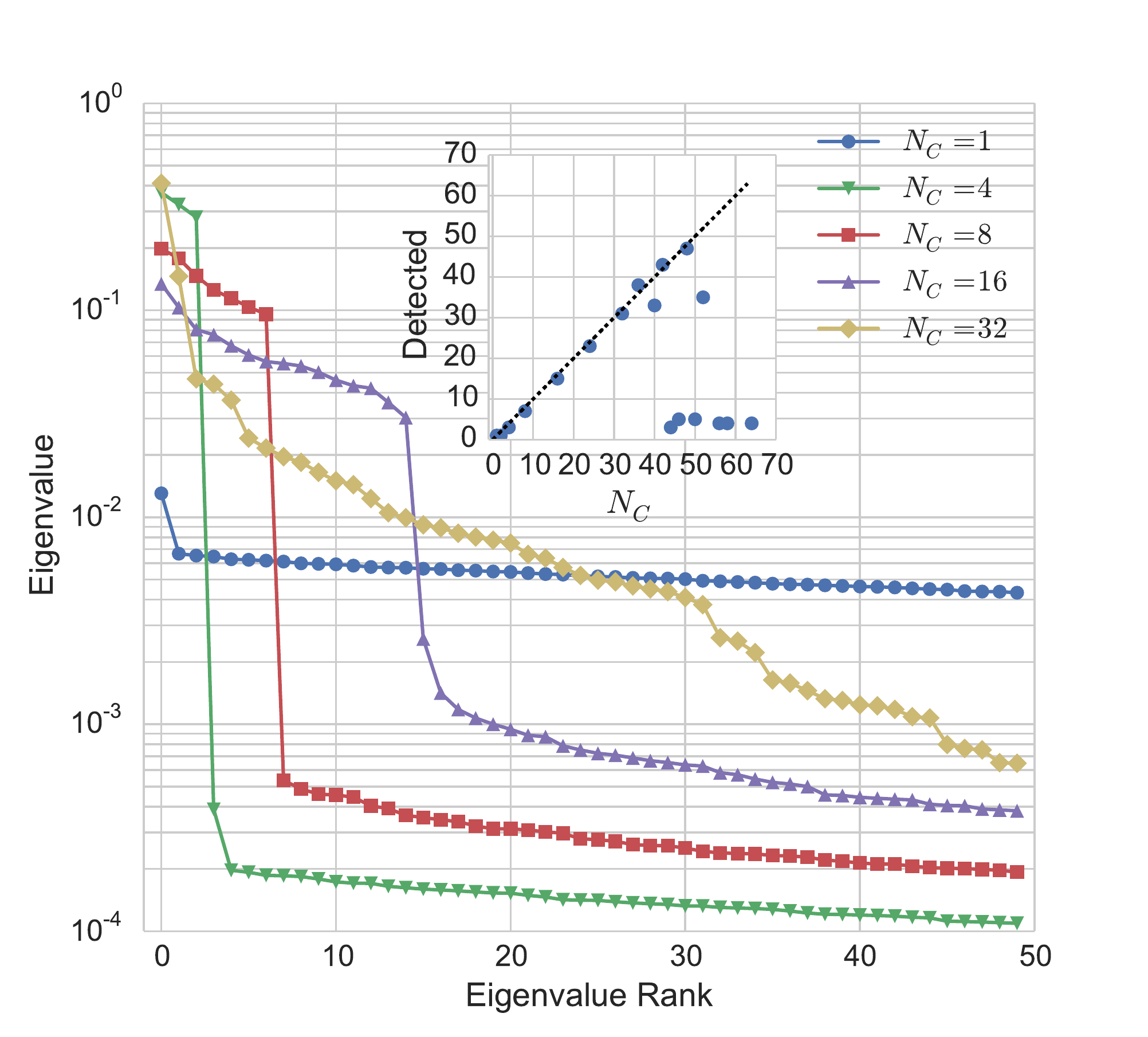}
 \caption{\label{TransitionTestCase}{\bf Eigenvalue plot corresponding to transition matrices with different numbers of clusters.} When the number of clusters is small, there is a clearly-defined gap between eigenvalues corresponding to cluster identity and eigenvalues corresponding to the fluctuations. As the number of clusters increases, the gap closes and it becomes more difficult to detect the structure of the population. Inset: Number of clusters detected by the algorithm versus actual number of clusters.}
\end{figure}

We use networks with $1000$ nodes, $N_L = 200$, $r_0 = 1.1$, $\delta r = 0.03$, $\gamma = 0.999$, and a variable number of clusters. We perform 250 iterations per run over $1000$ runs and use the resulting population vectors as our data points for the heredity metric. The results are shown in Fig.~\ref{TransitionTestCase}.

For small numbers of clusters, our algorithm successfully detects the cluster count. However, as the number of clusters grows, between-cluster transitions become more accessible due to the increasing number of between-cluster links compared to within-cluster links. This results in the gap in the eigenvalue plot shrinking, which makes it harder to detect the correct number of clusters (it is somewhat equivalent to what happens with a large mutation rate in the previous test case). When the number of clusters is larger than $32$ or so, the gap seems to disappear entirely (though there is clearly still some kind of structure in the eigenvalue plot) --- the result is that our metric detects only the steep slope between the first few eigenvalues, and predicts a very small number of clusters compared with the actual amount.

\subsection*{GARD}

The GARD model\cite{segre2000compositional,markovitch2012excess} is a model which has been observed to have a form of compositional heredity that is significantly different than the type of heredity obtained via information-carrying polymers\cite{lancet2014quasispecies}. In GARD, the fundamental objects of the system are vesicles which can be made from combinations of different types of lipids. The vesicles grow via a catalytic process --- each lipid type has some affinity with each other lipid type (given by the random affinity matrix $\boldsymbol{ \beta }$), and thereby controls the rate at which new lipids are added to the vesicle. When the vesicle reaches sufficient size $N_{\textrm{max}}$, it divides, randomly distributing its lipids to two child-vesicles of size $N_{\textrm{max}}/2$. As such, there is replication built into the model, but that replication takes place (in general) at a very low fidelity. As such, one of the key questions of GARD is whether or not high-fidelity states will naturally emerge from the dynamics as a result of a form of selection predicated on the robustness of a dynamical attractor.

To detect these high-fidelity replicators, the authors look at the similarity of a vesicle to its parent and its children\cite{markovitch2012excess}. This is used to filter out a subset of states that qualify as sufficiently high fidelity to be of interest. These states are then clustered using K-means clustering, with the silhouette metric used to determine the optimal number of clusters. As such, it is an example of an independent measurement of heritable states that we can use for comparison. The distinction is that in principle our metric should be able to operate without explicit knowledge of the pattern of replication of individuals. 

We follow the implementation of the GARD model as given in \cite{markovitch2012excess}, using specific $\beta$ matrices provided by the authors (as the distribution of values in the matrix is important to the degree of heredity observed). However, rather than using the repeated growth and division of a single individual vesicle, we simulate a population of vesicles under a distribution of selection pressures and then only permit our metric to observe the system-averaged compositions. The idea here is to test our ability to use competitive exclusion to extract out the individual-level heredity from an analysis of the system that is not able to distinguish the boundaries between individuals. We use populations of size $P=1$, $P=10$, and $P=100$, and simulate for approximately $100$ generations (direct examination of the timeseries data shows that even for $P=1000$ the system has reached steady by $30$ generations, and so this length of time should generally be long enough to consistently reach steady state for smaller populations).

Selection pressure was added by specifying a particular target composition $\vec{t}$ and then multiplying the elements of the affinity matrix $\boldsymbol{ \beta }$ according to:

\begin{equation}
 \boldsymbol{ \beta }_{ij} \rightarrow (1+\sigma t_i)\boldsymbol{ \beta }_{ij}
\end{equation}

where $\sigma$ measures the strength of the selection pressure. We find that selection pressures in the range between $\sigma=5$ and $\sigma=50$ seem to be sufficient to change the dominant state without destroying the structure of the underlying dynamics (whereas for $\sigma=500$ the underlying states are clearly distorted). We use $\sigma=50$ in our simulations, as the larger we can safely make $\sigma$ the more easily we can sample rarer heritable states. 

\begin{figure}
 \includegraphics[width=\columnwidth]{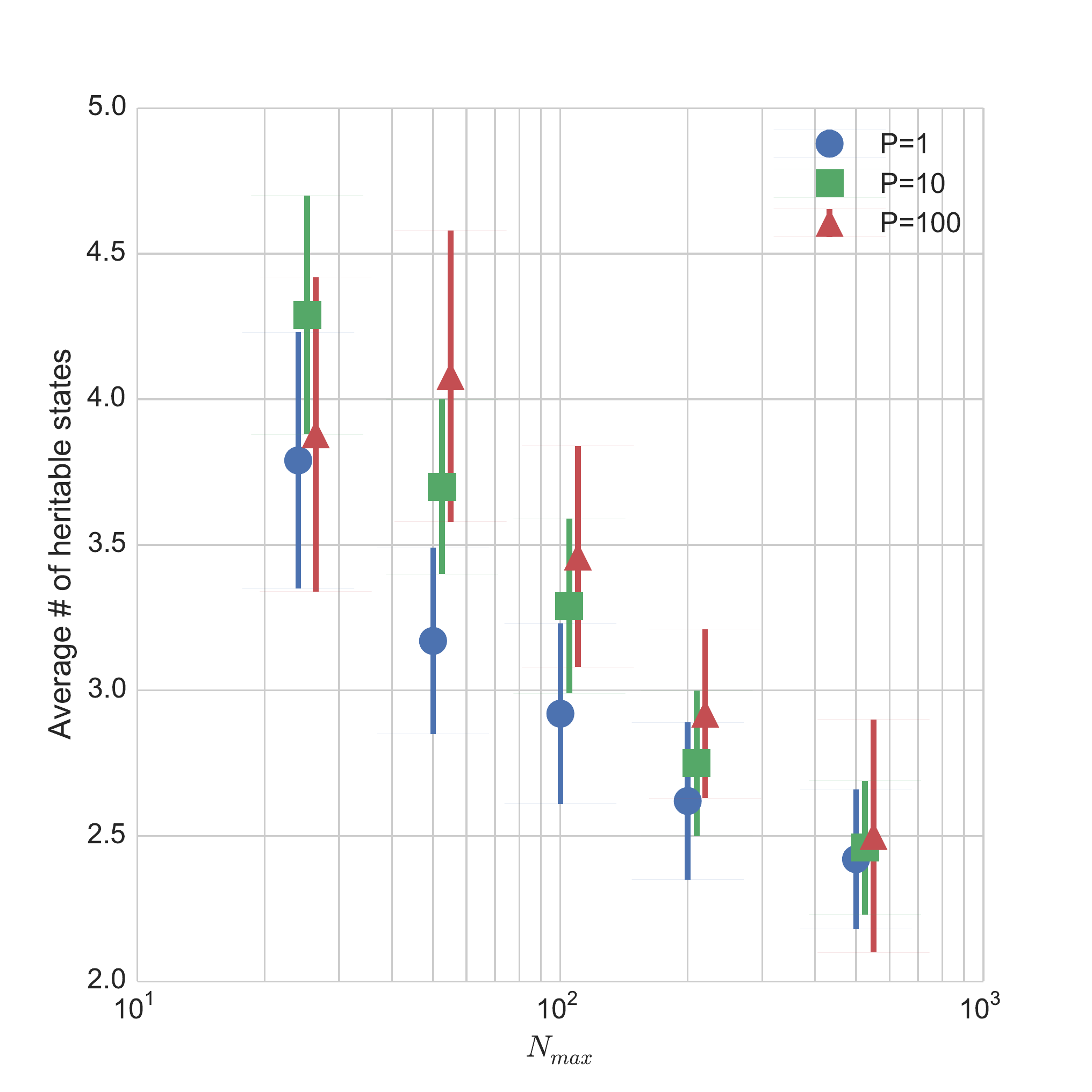}
 \caption{\label{GardSpecies}{\bf Hereditable states as a function of vesicle size.} In the GARD paper, the authors observed the trend that as the vesicle size $N_{\textrm{max}}$ was increased, the number of heritable states decreased\cite{markovitch2012excess}. We show that our algorithm can detect this trend successfully by plotting the average number of heritable states measured as a function of $N_{\textrm{max}}$. The points are slightly offset horizontally for clarity, but all are measured at the same values of $N_{\textrm{max}}$: $24,50,100,200,500$}
\end{figure}

We measure the mean number of heritable states for populations of size $1$, $10$, and $100$ averaged over $24$ different random $\boldsymbol{\beta}$ matrices. The measured results appear to be consistent with eachother to within the measurement errorbars. We find that the number of heritable states detected is consistent with the numbers previously reported --- on the order of about 3 distinct heritable states on average. Furthermore, we are able to detect the effect observed by the authors that the number of heritable states tends to decrease with the ratio of the vesicle size and number of distinct lipid types.  The results of our simulations are plotted in Fig.~\ref{GardSpecies}. 

\begin{figure}
 \includegraphics[width=\columnwidth]{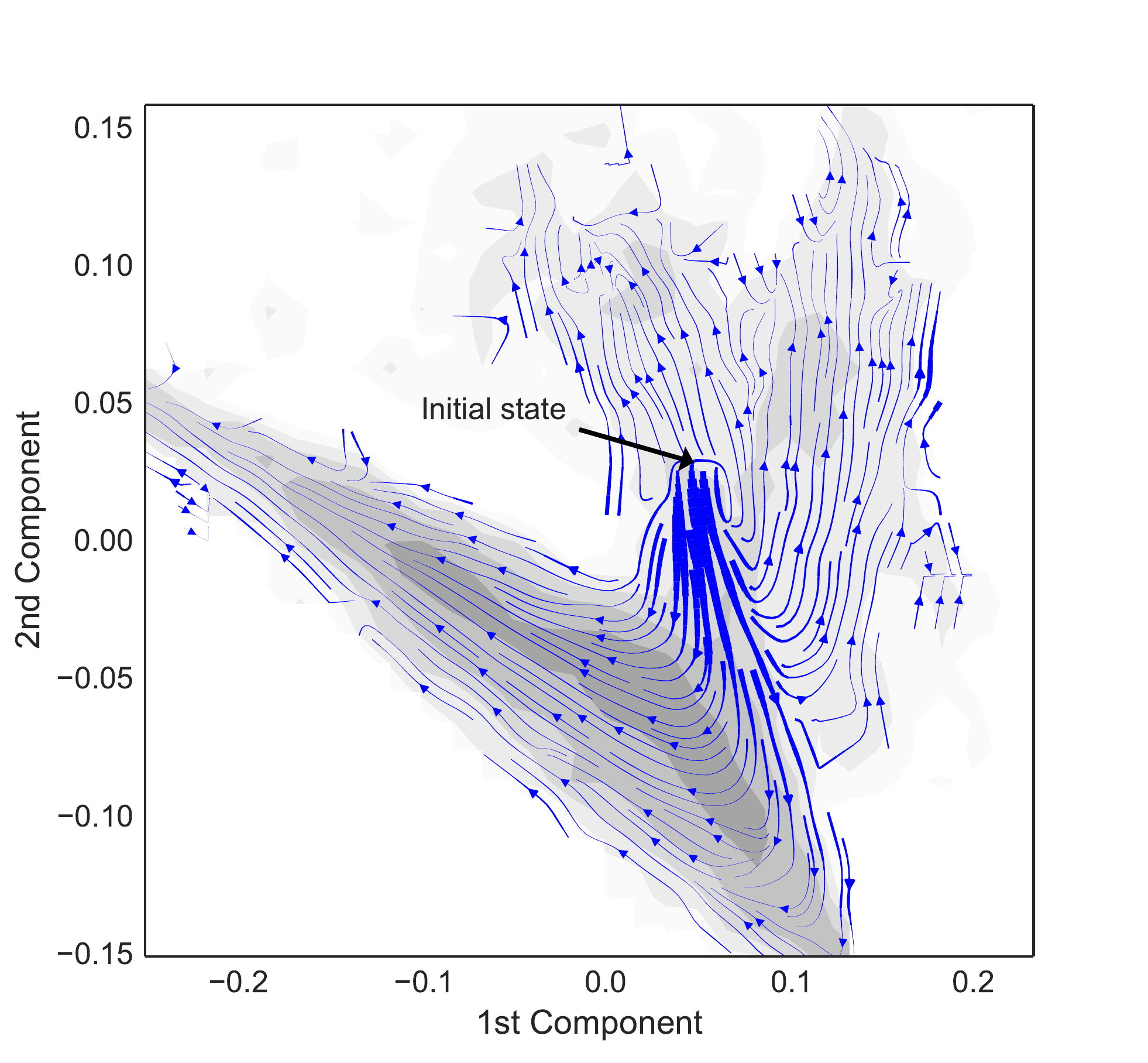}
 \caption{\label{GardFlowDiagram}{\bf Flow diagram of GARD dynamics in heritable-state space.} This figure shows the average time-dependent behavior of the system state for a particular choice of the $\boldsymbol{\beta}$ matrix and a distribution of selection pressures, projected onto the first two PCA eigenvectors. The thickness of the streamlines indicates the magnitude of the vectors in the underlying averaged vector field. As the mean vector field is the result of averaging a number of random walks, this can be thought of as the strength of the local bias. The underlying greyscale intensity shows the logarithm of the probability density for the system state being found at that location.}
\end{figure}

In order to check our intuition as to what the PCA metric is measuring and verify that it is detecting real distinctions, we look in detail at the dynamics of the GARD model for a specific choice of $\boldsymbol{\beta}$ for which our algorithm detected the existence of three species when $N_{\textrm{max}}=50$. We project the trajectory of the global system state  for a large population ($P=1000$) as a function of time onto the the first two eigenvectors of the PCA and then measure the mean direction of flow (averaged over the distribution of selection pressures) on the resulting 2D space. We find that individual trajectories go to isolated points in the space, but as we vary the selection pressure an overall structure emerges in which there are clearly multiple distinct regions of the phase space --- which region the system eventually ends up in depends on the particular selection pressure. One region corresponds to an extended linear subspace, where the details of the selection pressure position the system state at different places along the line. Most of the time the winning state is somewhere on this line. The other region is more broadly distributed but appears to have a second saddle point and a corresponding pair of distinct peaks. This structure is shown in Fig.~\ref{GardFlowDiagram}.

\section*{Discussion}

One of the difficulties of synthesizing the results of abstract modelling efforts is that often the details of the model are sufficiently idiosyncratic that it is difficult to directly compare the results of one model to another, or to experiment. A model that works in terms of a synthetic artificial chemistry may not easily make direct predictions about steady-state chemical concentrations in a real chemical system, because the details of chemical energies and reaction rates will likely be different. This means that although some insight can be gained, it is hard to say concretely what the one system shows us about the other. One way to overcome this limitation is to devise measurements which integrate out the details of the system but capture features that are somewhat representation-invariant. One could, for example, show that increasing the temperature in the synthetic system causes a certain statistical characteristic of the distribution of chemical compounds to change, and then ask whether or not the same trend is observed experimentally.

Much of the work being done to understand the emergence of life from prebiotic chemistry faces this difficulty. There are a number of different scenarios for how life could have emerged, and they each use different fundamental 'objects' to build models: the metabolism-first hypothesis uses chemical distributions in which heredity would be encoded in the particular pattern of attractors of the chemical system, the RNA-world hypothesis uses populations of information-containing polymers, and the lipid world hypothesis uses compositional information combined with cells that concretely identify the 'individuals' to which a given composition belongs. Experimental work on artificial cells usually uses some combination of these various features. Beyond that, there are a number of even more abstract computational models of artificial life which use anything from patterns embedded on a grid (e.g. cellular automata) to molecular dynamics simulations in which positional information might be relevant. If we want to understand the emergence of heredity and ask questions such as 'how does heritability scale with respect to various parameters?' in a transferrable way, we need a way to measure heritability which can apply to all of these systems.

We have presented a PCA-based heredity metric that helps bridge this gap. By applying a distribution of selection pressures, we can detect the number of heritable states of models even in the case in which competitive exclusion is allowed to proceed to completion. This makes it possible to analyze heritability and evolvability as intrinsic characteristics of a given model, rather than being tied to a particular length of observation of set of initial conditions. Furthermore, the metric we have introduced here is sufficiently general that it should be possible to apply it to a wide array of different abstract models as well as experimental data. Although the distribution of selection pressures and choice of relevant features must be customized appropriately for each model, the analysis itself is fairly simple to compute. 

One question that remains is whether or not it is possible to detect the distinction between limited and unlimited heredity using approaches of this nature. A modern biological organism has what can be thought of as unlimited heredity --- it is possible for the organism's genome to become longer if there is a need to store more information, so in practice the number of possible heritable states is infinite. On the other hand, the systems we have looked at in this paper have an intrinsic and fixed number of heritable states, and no obvious mechanism by which this number can be changed at will. We know from looking at the convergence properties of our metric that it tends to fail when the number of possible species is large compared to the number of data rows and features --- if we are dealing with a system that has a truly infinite number of heritable states, the likely outcome is that this metric will predict that there is only a single heritable state. If we are interested in discovering a system in which unlimited heredity exists or emerges, this is a significant problem. 

We have suggested two approaches to solving this problem. One approach is to not probe the overall heritability potential of the system as a whole, but instead to probe the heritability potential with respect to very specific types of selection pressure and then to combine these results into an overall picture of the evolvability of the system. The other approach is to reduce the scale of the system in order to reduce the combinatoric explosion of heritable states, and then measure how the number of heritable states scales with respect to system size. Heredity originating from replicating polymers is likely to scale differently than compositional heredity or heredity emerging from catalytic networks, and so this analysis may allow one to characterize different mechanisms of heredity as belonging to distinctive classes.

\section*{Acknowledgments}
We would like to acknowledge Piet Hut and Nathaniel Virgo for helpful discussions of our heredity metric, its meaning, and applications. We would also like to acknowledge Omer Markovitch for providing us with simulation parameters and data related to the GARD model.

\bibliography{references}

\end{document}